\documentclass[letter]{aa}  

\usepackage{graphicx}
\usepackage{txfonts}
\usepackage{xspace}
\usepackage{mathrsfs}
\usepackage{amssymb}
\usepackage{gensymb}

\newcommand{\nnhp}{$\rm N_2H^+$\xspace}
\newcommand{\nqnhp}{$\rm N^{15}NH^+$\xspace}

\newcommand{\nlow}{$\rm ^{15}N$\xspace}

\newcommand{\kms}{$\rm km \, s^{-1}$\xspace}

\newcommand{\tex}{$T_\mathrm{ex}$\xspace}
\newcommand{\vlsr}{$V_\mathrm{lsr}$\xspace}

\newcommand{\sigmav}{$\sigma_\mathrm{V}$\xspace}

\newcommand{\am}{$\rm NH_3$\xspace}
\newcommand{\qam}{$\rm ^{15}NH_3$\xspace}
\newcommand{\ratio}{$\rm ^{14}N / ^{15}N$\xspace}

\begin{document}

 \title{Nitrogen fractionation in ammonia and its insights on  nitrogen chemistry}

\titlerunning{First \ratio ammonia map}

   \author{E. Redaelli
          \inst{1}
    	\and
        L. Bizzocchi \inst{2}
        \and
        P. Caselli \inst{1}
        \and 
        {J. E. Pineda} \inst{1}
               }

   \institute{Centre for Astrochemical Studies, Max-Planck-Institut f\"ur extraterrestrische Physik, Gie{\ss}enbachstra{\ss}e 1, 85749 Garching bei M\"unchen,
Germany\\
              \email{eredaelli@mpe.mpg.de}
         \and
       Dipartimento di Chimica "Giacomo Ciamician", Universit\'a di Bologna, via F. Selmi 2, 40126, Bologna, Italy        }

   \date{...;...}

 
  \abstract
   {Observations of the nitrogen isotopic ratio \ratio in the interstellar medium are becoming more frequent thanks to the increased telescope capabilities. However, interpreting these data is still puzzling. In particular, measurements of \ratio in diazenylium revealed high levels of anti-fractionation in cold cores, which is challenging to explain.}
   {\cite{Furuya18}, using astrophysical simulations coupled with a gas-grain chemical code, concluded that the \nlow-depletion in prestellar cores could be inherited from the initial stages, when $^{14} \rm N ^{15}N$ is selectively photodissociated and $\rm ^{15}N$ atoms deplete onto the dust grain, forming ammonia ices. We aim to test this hypothesis. }
   {We targeted three sources (the prestellar core L1544, the protostellar envelope IRAS4A, and the shocked region L1157-B1) with distinct degrees of desorption or sputtering of the ammonia ices. We observed the ammonia isotopologues with the Green Bank Telescope, and we inferred the ammonia \ratio via a spectral fitting of the observed inversion transitions.}
   {\qam (1,1) is detected in L1544 and IRAS4A, whilst only upper limits are deduced in L1157-B1. The \am isotopic ratio is significantly lower towards the protostar {($\text{\ratio} = 210\pm 50$)} than at the centre of L1544 {($\text{\ratio} = 390\pm 40$)}, where it is consistent with the elemental value. We also present the first spatially resolved map of \am nitrogen isotopic ratio towards L1544. }
   {Our results are in agreement with the hypothesis that ammonia ices are enriched in \nlow, leading to a decrease of the \ratio ratio when the ices are sublimated back into the gas phase for instance due to the temperature rise in protostellar envelopes. The ammonia \ratio value at the centre of L1544 is a factor of 2 lower than that of \nnhp, which can be explained if a significant fraction of nitrogen remains in atomic form and if the ammonia formed on the dust grains is released in the gas phase via non-thermal desorption.}


   \maketitle
%

\section{Introduction}
Nitrogen (N) isotopic ratio \ratio is considered an important diagnostic tool to follow the evolution from the primitive Solar Nebula to the present time. In the Solar System, \ratio values range from values of $\approx 440-450$ in the Jovian atmosphere and in the Solar wind \citep{Fouchet04,Marty11}, which are believed to be representative of the Protosolar Nebula, to  $\mathrm{^{14}N/^{15}N} = 272$ in the Earth atmosphere \citep{Nier50}, down to isotopic ratios as low as $50$ in local spots of carbonaceous chondrites \citep{Bonal10}. This evidence is likely indicative of the presence of multiple nitrogen reservoirs at the moment of the formation of the planetary system \citep[cf.][]{HilyBlant17, Grewal21} 
\par
In the interstellar medium (ISM), measurements of the \ratio ratio are spread depending both on the kind of source and on the targeted molecular tracer. Nitriles  are usually enriched in $^{15}\rm N$ with respect to the elemental value of 400, with \ratio values of $140-360$  and of $160-460$ in low mass prestellar and protostellar cores, respectively \citep{HilyBlant13a, Wampfler14}. \cite{Spezzano22} reported the first \ratio map in HCN in a prestellar core, and their results show the importance of selective photodissociation in driving the nitrogen isotopic ratio of nitriles. 
\par
In low-mass prestellar sources, \nnhp appears instead deficient in $^{15}$N, with isotopic ratios of  $\mathrm{^{14}N/^{15}N} = 580-1000$ \citep{Bizzocchi13, Redaelli18}. A possible explanation for these pieces of evidence was proposed by \cite{Loison19}, invoking an isotope dependency of the reaction rate for dissociative recombination (DR) of diazenylium. Due to this, in the cold {($T \lesssim 10\, \rm K$) and dense ($n \gtrsim 10^4 \, \rm cm^{-3}$)} gas, where CO is depleted and DR is the main destruction route for diazenylium, \nqnhp is destroyed faster than the main isotopologue, leading to high \ratio values. This theory found partial confirmation in the work of \cite{Redaelli20}, where the authors showed that $\rm \text{\nnhp/\nqnhp}$ ratios in a sample of protostellar sources are significantly lower than in prestellar cores, suggesting that once CO dominates the destruction of \nnhp, its nitrogen isotopic ratio decreases back to the elemental value of $\approx 400$. 
\par
\cite{Furuya18}, on the other hand, proposed that the $^{15} \rm N $ antifractionation of diazenylium is inherited from the initial stages of the contraction when UV photons can still penetrate the cloud. At that time, N$_2$ is selectively photodissociated, leading to a \nlow enrichment in the atomic nitrogen (N\textsc{i}) gas. As the temperature decreases and density increases, N\textsc{i} freezes out onto the dust grains, where it rapidly reacts with hydrogen and leads to the formation of highly \nlow enriched ammonia ices. On the other hand, the adsorbed N$_2$ does not significantly react. As long as N$_2$ non-thermal desorption rate is higher than the \am one, the net effect is that the bulk gas results depleted in heavy nitrogen, while the \am ices are enriched. The direct consequence of this model is that, once the ammonia ices are evaporated back into the gas phase, one would expect to observe a lower $\mathrm{^{14}NH_3/^{15}NH_3}$ ratio with respect to the prestellar phases. 
\par
To the best of our knowledge, in the local ISM \qam detection has been reported only in Barnard 1 and NGC1333 \citep{Lis10}\footnote{\citet{Chen21} also reported \qam (1,1) detections in Barnard-1b and NGC1333, and the derived isotopic ratios are consistent within errorbars with those of \citet{Lis10}.}, with $\text{\ratio} = 334 \pm 50$ (Barnard 1) and $344 \pm 173$ (NGC1333), which are consistent with the nitrogen isotopic ratio in the local interstellar medium \citep[$300-400$;][]{HilyBlant17, Kahane18, Colzi18b}. \cite{Gerin09} reported a small sample of \ratio measurements in deuterated ammonia ($\rm NH_2D$). Their results span the range of $270-800$, but the double fractionation makes them more difficult to interpret, due also to the limited signal-to-noise (S/N) ratio. In this letter, we report the detection of \qam towards the prestellar core L1544 and the protostellar core NGC1333-IRAS4A (hereafter IRAS4A), and the non-detection in the shocked region L1157-B1. We also show the first map of \ratio in ammonia, obtained in L1544.

\section{Source selection and Observations}
The sources targeted in this work were selected to span a range of physical conditions. L1544 is a cold and highly concentrated prestellar core, reaching at its centre $T\sim 7\, \rm K$ and $n \sim 10^7 \, \rm cm^{-3}$\citep{Crapsi07, KetoCaselli15}. It shows already signs of gravitational contraction well extended also in its envelope \citep{Crapsi05,Redaelli22}. The uneven illumination
from the interstellar radiation field might be responsible for its diversified chemistry \citep{Spezzano17}. IRAS4A  is a Class 0 young stellar object, powering collimated molecular outflows \citep[cf.][]{Lefloch98}. Based on the detection of several complex organic molecules, it has been classified as a hot corino \citep{Bottinelli04}, where ---as a consequence of the high temperatures ($\sim 100 \, \rm K$)--- the ice mantles on the dust grains have begun to evaporate. L1157-B1 is the shock region associated with the bipolar outflows driven by the class 0 protostar IRAS 20386+6751. The shock causes the sputtering of the dust grains, releasing refractory and volatile materials back into the gas phase, leading to a rich chemistry \citep{Bachiller01, Arce08, Fontani14, Codella17}.

The observations were performed at the Green Bank Telescope (GBT) in Green Bank, West Virginia, in two runs in March 2019, using the K-band Focal Plane Array (KFPA) receiver, in combination with the VErsatile GBT Astronomical Spectrometer (VEGAS) backend. We used the observing mode 20, which allows combining the 8 VEGAS spectrometers with the 7 KFPA beams, using frequency switching (throw: $4.11\, \rm MHz$). The achieved spectral resolution is $5.9\, \rm kHz$ ($\approx 0.08\,$\kms at $23\, \rm GHz$). Four subbands were centred on the lines of interest: \am (1,1) at $23.694\, \rm GHz$, \am (2,2) at $23.722\, \rm GHz$, \qam (1,1) at $22.624\, \rm GHz$, and \qam (2,2) at $22.650\, \rm GHz$. \par
The data were acquired as small daisy maps with {radius $2'$ (first run) and $0'.75$ (second run)}. Mars and Venus were used as flux calibrators, and nodding observations (rotating among the seven beams) were used also to check the relative gains of the different beams. The data reduction was performed with the standard GBT KFPA pipeline \citep{Masters11}. The beam and aperture efficiencies were set at 0.95 and 0.69, respectively, in order to calibrate the data in the main-beam temperature scale. The imaging of the data was performed with \textsc{gbtgridder}\footnote{Publicly available at \url{https://github.com/GreenBankObservatory/gbtgridder}}. The data were gridded to a final pixel size of $6''$. At this stage, spectral smoothing was applied, downgrading the spectral resolution to $\approx 0.15\, $\kms, in order to improve the signal-to-noise ratio (S/N). The final beam size is $\approx 36''$.

The noise levels reached in the $0.15\,$\kms channel for the main isotopologues are $rms = 20\, \rm mK$ (L1544), $40\, \rm mK$ (IRAS4A), and $30\, \rm mK$ (L1157-B1). For the \qam (1,1) line the datacubes present typical averages of $rms \approx 30 \, \rm mK$. The \qam (2,2) 
 line is not detected in any source. The integrated intensity maps for the main isotopologues, and for \qam towards L1544 are presented in Appendix~\ref{app:Moment0}. In this work, we focus only on the central region (radius: $120''$) of each source, where the noise in the datacubes is approximately constant. 

\section{Analysis}
Our main goal is to infer the nitrogen isotopic ratio \ratio from the column densities of the two ammonia isotopologues. These are derived by performing a spectral fit to the observed spectra, which is described in detail in Appendix~\ref{app:SpectralFit}. 
In the following subsections, we describe the results for each source.

\subsection{L1544 \label{sec:L1544}}
\begin{figure*}[!h]
\centering
\includegraphics[width=\textwidth]{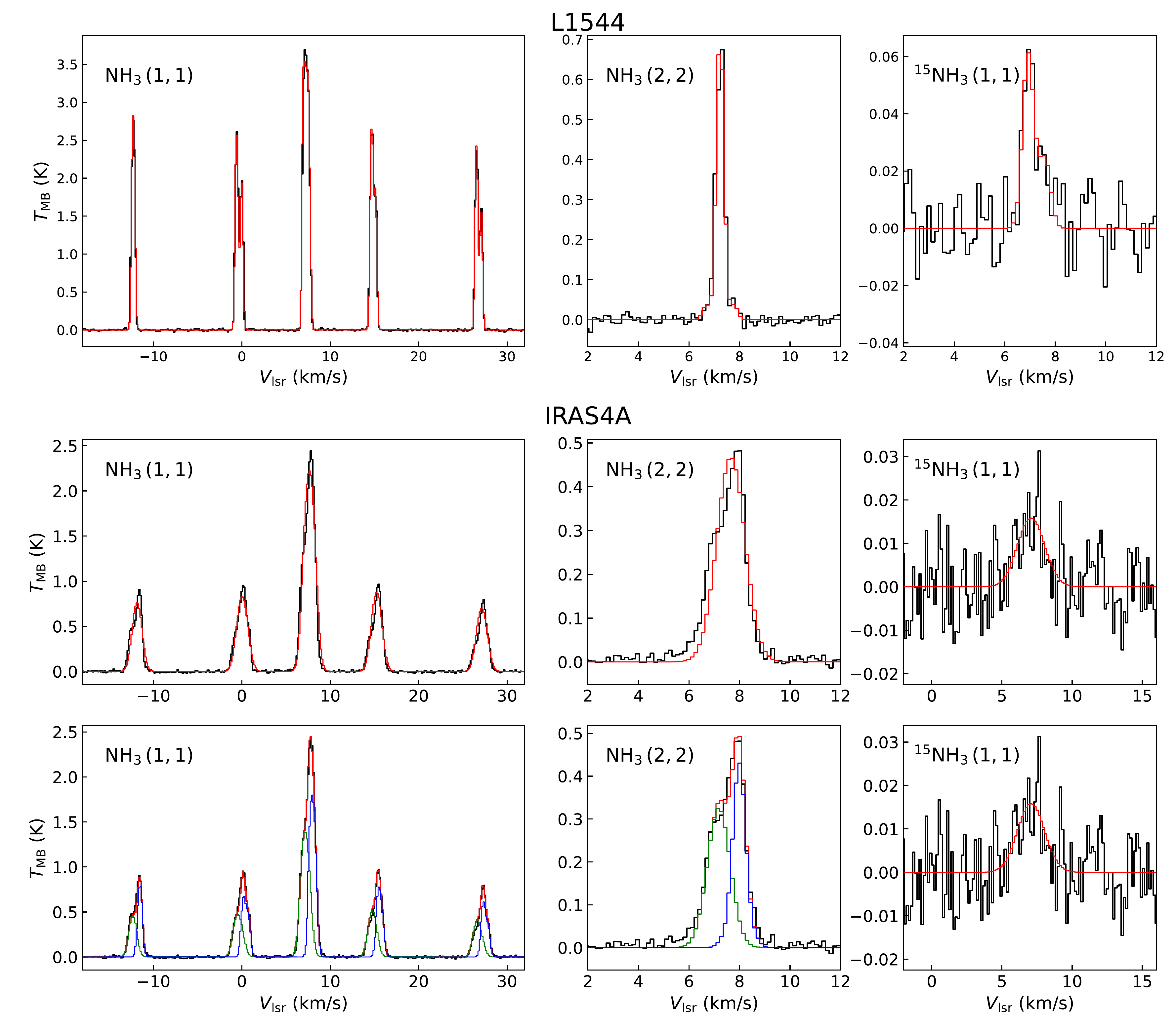}
\caption{Top row: results of the fit to the spectra extracted in the central beam, in L1544. The observed spectra are shown with black histograms, whilst the best-fit models are presented in red. For \am (2,2), we only show the central main hyperfine group. Central panels: same as the top row, but this time for IRAS4A. The spectra are averages over the central 90''. The best-fit models obtained using a single velocity component approach are shown with red histograms. Bottom panels: the left and central panels show the results of the two-component fit performed on the \am lines in IRAS4A. The two velocity components are shown in green (warm component) and blue (cold component), whilst the total spectrum is shown in red. In the right panel, the red curve shows the one-component fit to the \qam line, using the warm component fit results as input for the temperature values.\label{fig:centSpec}}.
\end{figure*}

We extract the spectra of all detected transitions in the central beam ($36''$) around the millimetre dust peak, in order to improve the $\rm S/N$ of the \qam (1,1) line. The resulting data, together with the best fit obtained with \textsc{pyspeckit}, are shown in the top row of Fig.~\ref{fig:centSpec}, and the best-fit solution is presented in Table \ref{tab:fit_results}. The \vlsr of the two species are consistent within uncertainties, which suggests that they arise from the same region in the source. However, it is important to notice that the \sigmav of \qam is larger than, and marginally inconsistent with, that of the main isotopologue. With the column density values derived, we obtain an isotopic ratio of $\text{\ratio} = 390 \pm 40$.

%

\begin{table*}[h]
    \renewcommand{\arraystretch}{1.4}
\caption{Best-fit solutions obtained for \am and \qam lines towards L1544 and IRAS4A. For the latter, two sets of values are reported, corresponding to the one-component and two-component approaches (see the main text for more details). \label{tab:fit_results}}
\centering
\begin{tabular}{cccccc}
\hline 
Species &	$T_\mathrm{rot}$ & \tex & $\log_{10}  \left [ N_\mathrm{col} \right ] $ & \sigmav & \vlsr \\ 
	&	(K)			&	(K)	&	 ($\log_{10}   [ \mathrm{cm}^{-2} ] $)& (\kms) & (\kms) \\
\hline \hline
\multicolumn{6}{c}{L1544} \\
\am & $8.74 \pm 0.06$    & $6.322\pm 0.016$   & $14.970 \pm 0.003$  & $0.1464 \pm 0.0006$   & $7.1603 \pm 0.0007$      \\ 
\qam & $8.74 $        & $6.322 $       & $12.38 \pm 0.05$      & $0.20 \pm 0.04$       & $7.20 \pm 0.04$          \\ 
\hline
\multicolumn{6}{c}{IRAS4A - 1 component} \\
\am & $12.55 \pm 0.16$   & $6.07 \pm 0.11$   & $14.715 \pm 0.012$  & $0.528 \pm 0.006$   & $7.585 \pm 0.005$   \\ 
\qam & $12.55$       & $6.07 $       & $12.31 \pm 0.08$      & $0.9 \pm 0.2$       & $7.2 \pm 0.2$       \\
\hline
\multicolumn{6}{c}{IRAS4A - 2 component{s}} \\
\am {warmer} component & $13.22 \pm 0.11$     & $5.08 \pm 0.07$     & $14.474 \pm 0.012$    & $0.423 \pm 0.008$     & $7.107 \pm 0.013$     \\ 
\am {colder} component & $11.97 \pm 0.09$     & $5.06 \pm 0.07$     & $14.522 \pm 0.006$    & $0.245 \pm 0.003$     & $7.915 \pm 0.004$     \\
\qam & $13.22 $       & $5.08$       & $12.39 \pm 0.08$      & $0.9 \pm 0.2$       & $7.2 \pm 0.2$       \\
\hline
\end{tabular}
\end{table*}

\subsection{IRAS4A}
The \qam (1,1) line is undetected with the obtained $rms$ level towards IRAS4A. In order to obtain a tentative detection at the $2\sigma$ level, we compute the average spectra in a central area of $45''$ of radius, shown in Fig.~\ref{fig:centSpec}. 
We initially perform a single-velocity-component spectral fit, shown in the middle row panels in Fig.~\ref{fig:centSpec}. The obtained best-fit parameters are summarised in Table~\ref{tab:fit_results}. Using the obtained values for the column densities of the two isotopologues, we compute the isotopic ratio of $\text{\ratio} = 260 \pm 50$. \par
The best-fit solution for the \am transitions is unable to reproduce the spectral feature seen in the observations and in particular the blue shoulder at low velocities (see Fig.~\ref{fig:centSpec}). This feature was also detected by \cite{Lis10}, and it is likely due to the blue lobe of the protostellar outflow, which enters the large GBT beam and the spatial average we performed. 
We have then modelled the \am lines using two velocity components, and these results are also reported in Fig.~\ref{fig:centSpec} and  Table~\ref{tab:fit_results}. The total \am emission is due to two contributions, {a warmer} component --- with $T_\mathrm{rot} = (13.22 \pm 0.11)\,$K--- at lower \vlsr, and {a colder} component, characterised by higher \vlsr and  a lower temperature, $T_\mathrm{rot} = (11.97 \pm 0.09)\,$K. {Although the temperature difference is only of $\sim 1.2\,$K,} the velocity dispersion of the {colder} component is almost half that of the {warmer} one, supporting the scenario that its emission arises in the colder and more quiescent gas surrounding the hot corino. {The \vlsr of the warmer component ($7.1\,$\kms) is consistent with high-resolution line observations of the hot-corino (see e.g. \citealt{Bottinelli04,Choi11,Yamato22}, who all reported systemic velocities of $6-7\,$\kms), further supporting the idea that this component arises from the warmer gas closer to the central protostar.} The improvement of the fit using two velocity components has been confirmed with the Akaike information criterion ({as done e.g. by \citealt{Choudhury21};} see Appendix~\ref{app:AIC_analysis}).
\par 
The low $\rm S/N$ \qam (1,1) spectrum prevents fitting two velocity components. The best-fit solution for the one-component fit  is found at $V_\mathrm{lsr}  = (7.2 \pm 0.2)\, $\kms, which is inconsistent with the result of the single-component fit to the \am lines, and it is instead consistent with the velocity of the warm component (cf. Table~\ref{tab:fit_results}). The \qam (1,1) line appears significantly broader than both velocity components, but its linewidth is closer to that of the warmer component. We can speculate, hence, that the emission of the rarer isotopologues arises mostly from the warm gas closer to the central object. We have re-fitted the \qam (1,1) line, using the temperatures of the warm component as fixed parameters. The nitrogen isotopic ratio for the warm component only is $\text{\ratio} = 210 \pm 50$.


\subsection{L1157-B1}
The \qam (1,1) line is undetected at the shock position L1157-B1. The main isotopologue lines present broad and asymmetric line profiles, with extended blue wings, due to the presence of the outflow and the shock. In order to estimate a lower limit on the nitrogen isotopic ratio, we have used the results of \cite{Umemoto99}, who reported the detection of the first six inversion transition in the source and estimated the column density using the rotational diagram method\footnote{The strong asymmetry in the spectra prevent us from using the LTE fit implemented in \textsc{pyspeckit}.}. 
The noise level in the \qam spectrum extracted from the central $72''$ (equal to the beam size of the Nobeyama telescope used in \citealt{Umemoto99}) is $rms =7\, \rm mK $.  Using the optically thin approximation (see Appendix A of \citealt{Caselli02b}), and assuming a line width of $FWHM=1 \, $\kms, we obtain an upper limit of $N \rm _{col} (^{15}NH_3) \leq 6 \times 10^{11} \, cm^{-2}$ and a lower limit on the isotopic ratio of $\text{\ratio} \geq 160$.

\subsection{\ratio map in L1544\label{sec:L1544_map}}
The \qam line towards L1544 is bright enough to infer a map of the nitrogen isotopic ratio in ammonia, the first of its kind in an astrophysical source to our knowledge. The map, shown in the central panel of Fig.~\ref{fig:L1544_map}, is derived following the procedure described in Appendix~\ref{app:SpectralFit}. Despite its small extension, some spatial trends can be seen. The isotopic ratio is higher at higher column densities. The average value for positions where $N\rm (H_2) > 2\times 10^{22} \, cm^{-2}$ is $\text{\ratio} = 390 \pm 60$, where the uncertainties are the standard deviation across the mean value. The ratio decreases at lower column densities, as visible in particular in the southeast direction, where its range is $150-250$. \par
 In order to confirm this spatial trend with higher S/N, we have computed the average spectra over one beam at three different offsets (also shown in Fig.~\ref{fig:L1544_map}). The first offset (labelled A) is centred in the  $N\rm (H_2)$ peak\footnote{The $N\rm (H_2) $ map is computed from \textit{Herschel} data, and its peak is $\approx 10'' $ shifted with respect to the millimetre dust peak, at which the spectra of Fig.~\ref{fig:centSpec} are taken.}. Offsets B and C are chosen to cover approximately the same  $N\rm (H_2)$ values in the southeast and northwest direction, respectively. The isotopic ratios obtained fitting the average \am and \qam spectra are $420 \pm 50$ in A, $200\pm 40$ in B, and $290\pm 70$ in C. The difference of the isotopic ratios between A and B is significant above the $3\sigma$ level ($220\pm60$), whilst that between A and C is instead detected only at $1.5\sigma$ level ($130 \pm 90$). Based on these results, we can certainly confirm a decrease of \ratio in the south-eastern direction.
 \par
 { The ammonia \ratio map is strikingly similar to that of HCN reported by \cite{Spezzano22} both regarding the morphology and the absolute values. Those authors also found a significant decrease of \ratio (although only in the south-eastern direction), from $\text{\ratio} = 440 \pm 60$ at the core's centre down to $220 \pm 30$ towards a position close to offset B in Fig.~\ref{fig:L1544_map}.}

\begin{figure*}[!h]
\centering
\includegraphics[width=\textwidth]{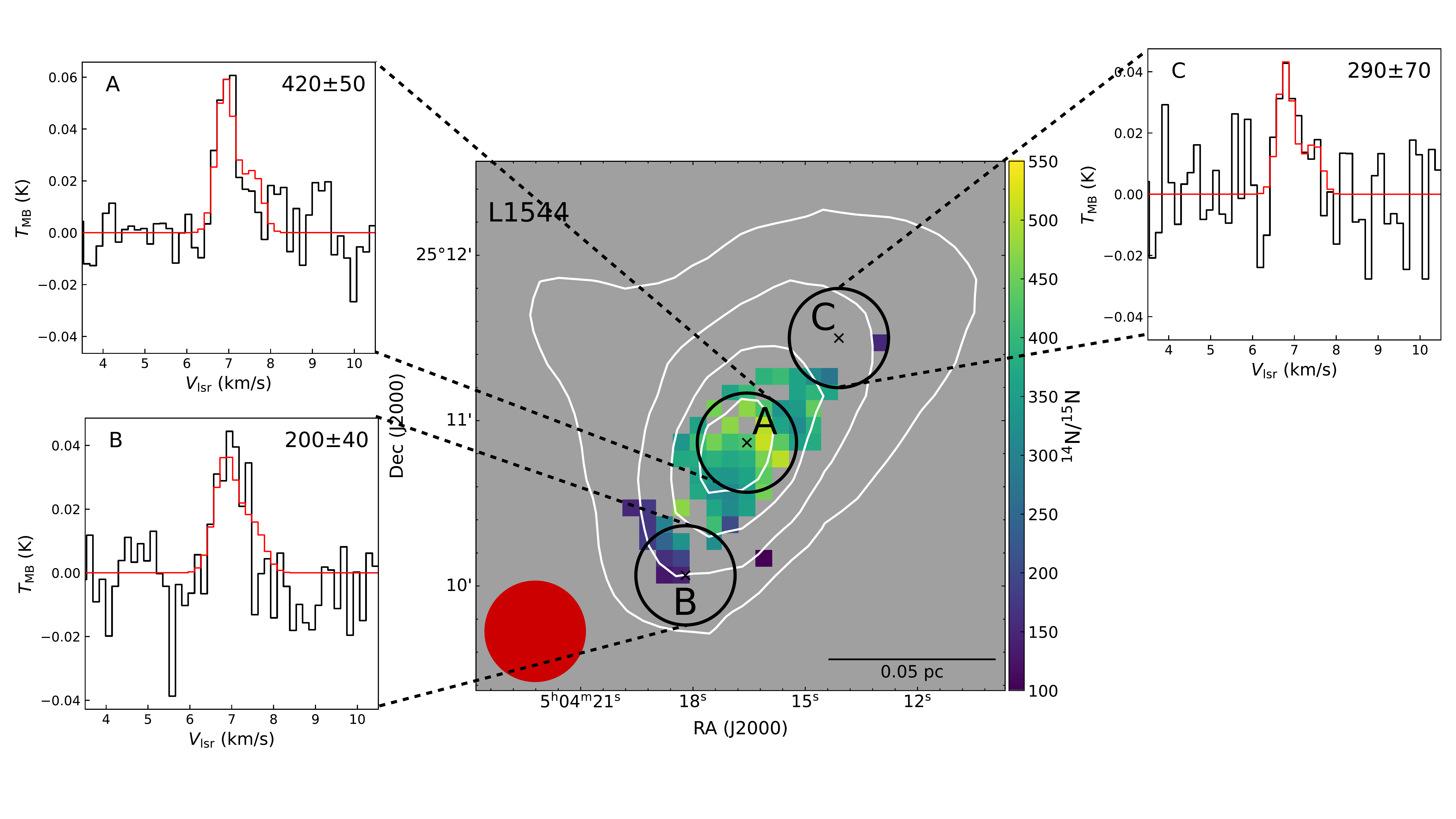}
\caption{ Central panel: ammonia \ratio map in L1544. The contours show the total gas density at levels $N\rm (H_2) = [1, 1.5, 2, 2.5]\times 10^{22} \, cm^{-2}$, obtained from \textit{Herschel} {SED fitting} \citep{Spezzano17}. The black circles show the three offsets (labelled with capital letters) where we compute average spectra to improve S/N. The averaged \qam (1,1) lines are shown in the surrounding panels (black: observations, red: best-fit models, obtained as described in App.~\ref{app:SpectralFit}). These panels are labelled in their top-left corner, whilst in the top-right corners we indicate the measured \ratio. \label{fig:L1544_map}}.
\end{figure*}

\section{Discussion and conclusions}
In this work, we have measured the ammonia nitrogen isotopic ratio \ratio towards three sources, a prestellar core (L1544), a Class 0 protostar (IRAS4A), and the shocked region L1157-B1. The first result is that at the centre of L1544, where the temperature is the lowest in the sample ($T_\mathrm{rot} \lesssim 9\, $K), the measured ratio is $390 \pm 40$, significantly higher than the value in IRAS4A, where ---using a two-component fit for the main isotopologue--- we obtained $\text{\ratio}=210 \pm 50$, which is consistent with the results of \citet{Lis10} (who pointed at a $\approx 20''$ offset position). The difference among the two sources is $\text{\ratio}_\mathrm{L1544} -\text{\ratio} _\mathrm{IRAS4A}  = 180 \pm 60$, significant at the $3\sigma$ level. In L1157-B1, \qam is undetected, yielding to a lower limit $\text{ratio} \geq 160$. These results are in agreement with the scenario where gas-phase ammonia present an isotopic ratio close to the elemental value of $\approx 400$ in cold and quiescent sources (L1544). The ammonia ices are instead enriched in \nlow, and this enrichment is detected in the gas phase when the icy mantles are either evaporated due to the increase of temperature or destroyed by shocks. These findings would hence support the theory of \cite{Furuya18}.
\par
This work provides the first spatially-resolved ammonia \ratio map. At the centre of the prestellar core L1544, the obtained \ratio is consistent both with the Protosolar Nebula value and with the elemental value in the local ISM. On the other hand, it is almost a factor of two lower than that of \nnhp, derived using single-dish data at a similar resolution ($\text{\nnhp}/ \text{\nqnhp} = 920^{+300}_{-200}$, \citealt{Redaelli18}). This can be explained if a significant fraction of nitrogen is in atomic form also in the high-density and low-temperature centre of prestellar cores, as suggested by \citet{Maret06}, and if \am can efficiently form on the surface of dust grains (via hydrogenation of atomic nitrogen), followed by reactive desorption. In fact, as \am formation in the gas phase starts from the reaction $\rm N_2 + He^+ \rightarrow N^+ +N+He$, and given that N$_2$ is \nlow-deficient toward the core centre (as measured by \nnhp), the same antifractionation should be found in \am. As this is not the case, atomic \nlow-enrichment is needed to allow the \nlow-fractionation of \am to reach the ISM values, i.e. about a factor of two higher than that of \nnhp. One possible explanation is that a significant fraction of the nitrogen is still in the atomic form in the core centre, maintaining a low \ratio value due to selective photodissociation of $\rm N^{15}N$. The \nlow enrichment is then transmitted to ammonia forming on the dust grains, from where it can be desorbed through reactive desorption. This is especially true toward the south of L1544, which is more exposed to the interstellar radiation field \citep[cf.][]{Spezzano16} favouring the selective photodissociation, where in fact we detect a decrease of \ratio. {The fact that ammonia mainly comes from the surface hydrogenation of atomic nitrogen would also explain the similarities between the \ratio map of \am and of HCN \citet[cf.][]{Spezzano22}, as nitriles are formed from atomic nitrogen also in the gas phase.} A further element to take into consideration is that at the core centre, the abundance of $\rm He^+$ increases due to the CO catastrophic freeze-out \citep[see e.g.][]{Sipila19}, thus making the gas-phase formation route more efficient and then decreasing the \nlow-fractionation. {Recent works revealed that ammonia isotopologues suffer from depletion at the high densities ($n> 10^5 \, \rm cm^{-3}$) characteristic of dense prestellar cores, which however can be seen only with high-resolution interferometric data (e.g. the core H-MM1 seen with VLA, \citealt{Pineda22}; or L1544 observed with ALMA, \citealt{Caselli22}). We exclude hence that this phenomenon is important at the scales probed by our GBT data.}
\par
{Another possible explanation to these observational evidence is that the \nnhp isotopic ratio is not tracing the bulk of N$_2$ gas, due to isotope-selective dissociative recombination \citep[as suggested by][]{Loison19}. The \ratio measurement in \nnhp is available only at the dust peak, however, preventing us to draw conclusions on the meaning of the \am isotopic ratio towards the south-east of the core in this scenario (beyond what already discussed). Laboratory results testing this hypothesis are needed to confirm if the nitrogen isotopic ratio of \nnhp is indeed indicative of that of molecular nitrogen, or if it experiences other fractionation processes.} 
\par
Our results show that significant efforts are still necessary to understand nitrogen chemistry. The derived \ratio values also show that non-thermal desorption mechanisms are important and must be taken into account in astrochemical models.

   \begin{acknowledgements}
   {The authors kindly thank the anonymous referee for the comments that helped improve the manuscript.} ER, PC, and {JEP} acknowledge the support of the Max Planck Society. ER thanks Dr. J. Chantzos, for the help and support during the data acquisition. 
  \end{acknowledgements}

 \bibliographystyle{aa}
 \bibliography{Literature}
 \newpage
\appendix

\section{Integrated intensity maps for the detected transitions \label{app:Moment0}}

\begin{figure}[!b]
\centering
\includegraphics[width=0.5\textwidth]{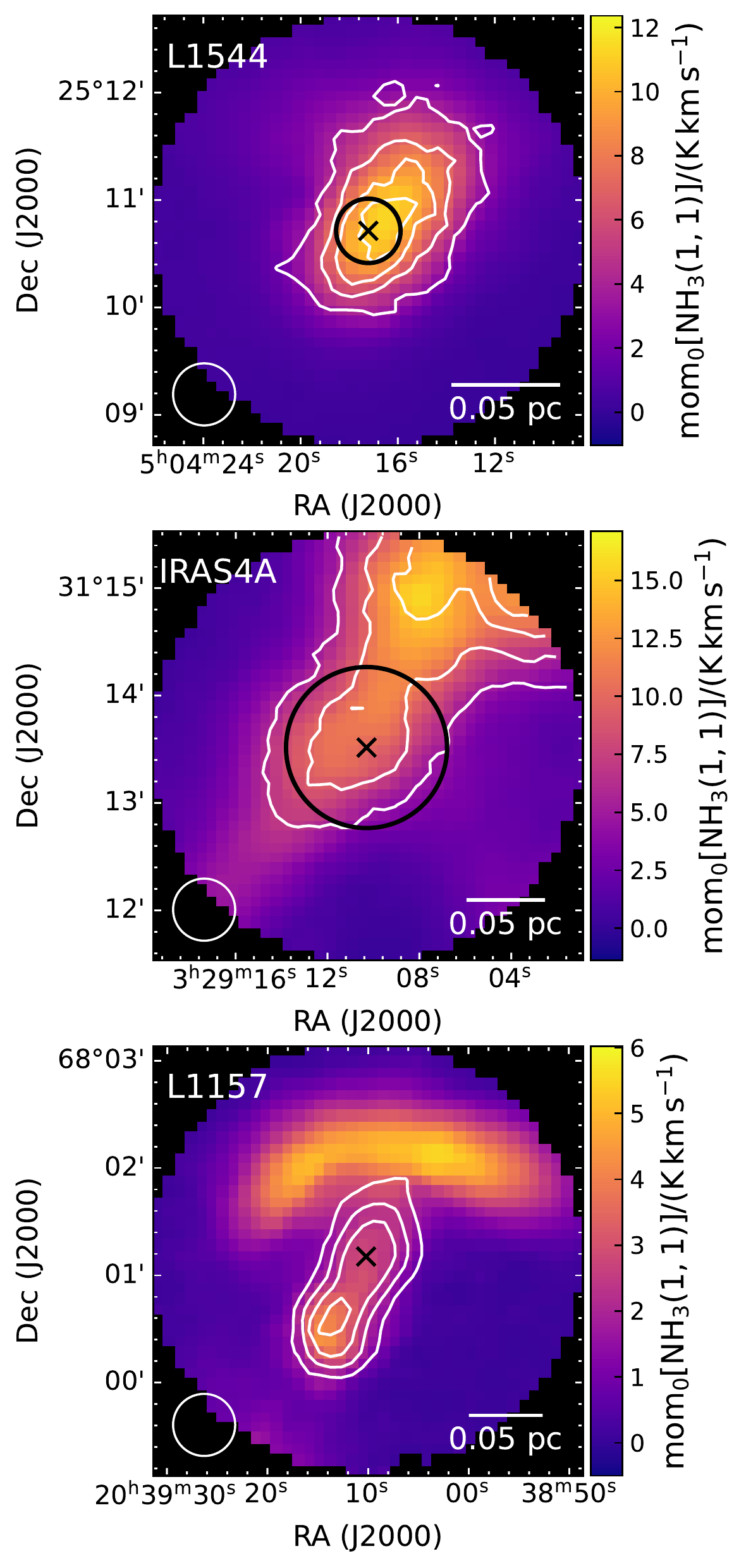}
  \caption{ The colorscale in each panel shows the integrated intensity of the \am (1,1) line. {The central pointings (see Main Text) are shown with a black cross.} The contours show the integrated intensity of the corresponding (2,2) transition, at levels of $0.3,0.5,0.7,0.9$ of the peak value, which is $0.4\rm \, K \, km \, s^{-1}$ (L1544, top panel), $1.6 \rm \, K \, km \, s^{-1}$ (IRAS4A, central panel), and $2.3 \rm \, K \, km \, s^{-1}$ (L1157-B1, bottom panel). The beam size and scale bar are shown in the bottom left and right corners, respectively, of each panel. {The black circles in the top and central panels show the areas where the spectra in Fig.~\ref{fig:centSpec} are extracted.}} \label{fig:NH3_mom0}
\end{figure}

Figure \ref{fig:NH3_mom0} present the maps of the integrated intensity of the main isotopologues transitions towards the three targets: L1544 (distance $d=170\, \rm pc$, \citealt{Galli19}; coordinates $\alpha_{\rm J2000} = \rm 05^h \,04^m \, 17^s.2$, $\delta_{\rm J2000} = \rm 25^\text{\degree} \,10^m\, 41^s.8$), IRAS4A (distance $d=235\, \rm pc$, \citealt{Hirota08}; coordinates $\alpha_{\rm J2000} =\rm 03^h \,29^m \, 10^s.3$, $\delta_{\rm J2000} = \rm 31^\text{\degree} \,13^m\, 31^s.0$), and L1157-B1 (distance $d=250\, \rm pc$, \citealt{Arce08}; coordinates $ \alpha_{\rm J2000} =\rm 20^h \,39^m \, 10^s.2$, $\delta_{\rm J2000} =\rm 68^\text{\degree} \,01^m\, 10^s.5$). These are computed by selecting the channels with line emission (i.e. considering all hyperfine components), and excluding channels with a noise level higher than $2\sigma$. Uncertainties on the integrated intensity are computed as $rms_\mathrm{II} = \sqrt{N_\mathrm{ch}}\times \delta V \times rms$, where $N_\mathrm{ch}$ is the number of channels over which the integration is performed, and $\delta V $ is the channel width. The resulting average uncertainties for the \am (1,1) and (2,2) lines are $16 \rm \, mK \, km \, s^{-1} $ and $6 \rm \, mK \, km \, s^{-1} $ for L1544, $42 \rm \, mK \, km \, s^{-1} $ and $20 \rm \, mK \, km \, s^{-1} $ for IRAS4A, and $26 \rm \, mK \, km \, s^{-1} $ and $15 \rm \, mK \, km \, s^{-1} $ in L1157-B1. \par
{ The \am emission in L1544 for both transitions shows a core-like structure,  roughly centred around the dust peak. The core has an elliptical shape, with the major axis in the northwest-southeast direction. Towards IRAS4A, the two \am lines show an elongated morphology, with a secondary peak close to the position on the hot corino. The intensity peaks, however, are found in the northeast limit of the FoV; this is consistent with larger scale maps of the region \citep[see e.g.][]{Friesen17}, which show a peak of \am intensity in that direction. Towards L1157, the \am (1,1) and (2,2) transitions present different morphology, due to the presence of the outflow and the associated shocks. The (1,1) integrated intensity presents a flattened structure in the north of the FoV, in correspondence with the dense core surrounding the protostar IRAS 20386+6751 \citep[cf.][]{Bachiller93}. A weaker emission is also seen along the blue-shifted outflow, towards the knots B1 (centre of the map in Fig.~\ref{fig:NH3_mom0}) and B2 (immediately to the southeast of B1). The integrated intensity map of \am (2,2) is instead dominated by the emission in the shocks along the outflow, which leads to increasing intensities and linewidths of this transition \citep[as shown also by][using the (1,1) and (3,3) transitions]{Tafalla95}.}
\par
\begin{figure}[!h]
   \centering
   \includegraphics[width=0.5\textwidth]{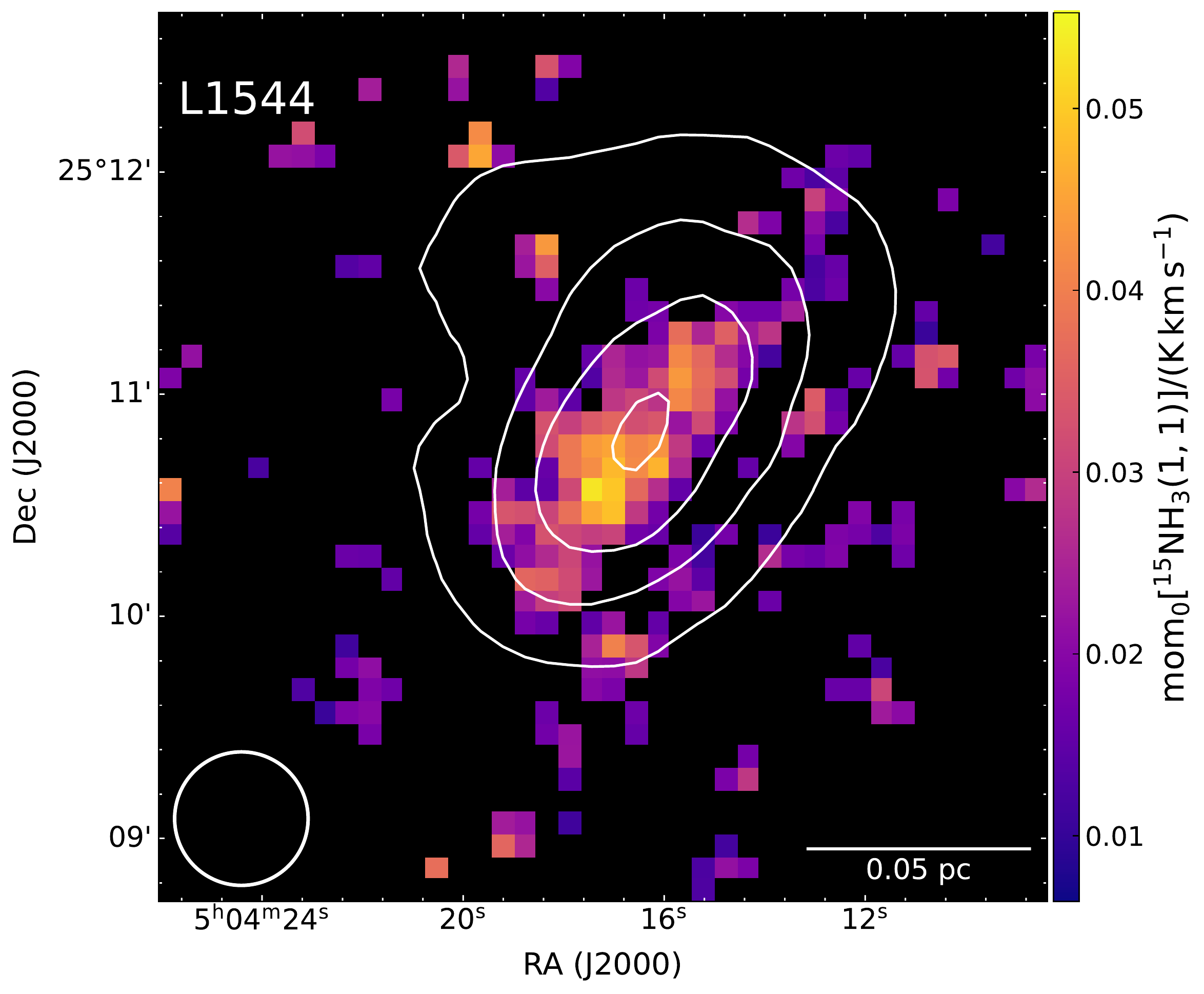}
      \caption{The integrated intensity of \qam (1,1) in L1544{, masked following $\rm S/N <3$}. The contours show the integrated intensity of the \am (1,1) line (the levels start at $2\, \rm K \,$\kms and increase with a step of  $3\, \rm K \,$\kms). The average noise level of the map is  $6\, \rm mK \,$\kms. The beam size and scale-bar are shown in the bottom left and right corners, respectively.}
         \label{fig:15NH3_L1544}
   \end{figure}

The \qam (1,1) is quite weak, reaching peak intensities of $\approx 60\, \rm mK$ at the centre of L1544, and of $\approx 20  \, \rm mK$ at the centre of IRAS4A. The transition is not detected in L1157-B1. In the case of L1544, the line is bright enough to map its integrated intensity, and the result is shown in Fig.~\ref{fig:15NH3_L1544}.

\section{Spectral fitting procedure\label{app:SpectralFit}}
The {spectral fit is performed using the \textsc{python} package \textsc{pyspeckit}, which already contains the full spectroscopic description of the \am inversion transition \citep{Ginsburg11,Ginsburg22}}. In order to perform a consistent analysis of both isotopologues, we have recently updated the package to include also the model for the \qam (1,1) and (2,2) lines. In particular, we have taken the frequencies, the rotational constants, and the Einstein coefficients for spontaneous emission values from the CDMS catalogue. We have performed the calculations for the hyperfine structures, as follows. The component position and intensities of the hyperfine structure of the $^{15}$NH$_3$  $J_K = 1_1-1_1$ and $2_2-2_2$  transitions have been computed from a complete reanalysis of the literature molecular beam data \citep{Kukolich67,Kukolich68,Hougen72}. Such high resolution ($\sim 0.3$\,kHz) measurements allowed the discrimination between subtle hyperfine effects and the determination of the coupling constants for $^{15}$N and H spin--rotation, as well as $^{15}$N--H, and H--H spin--spin coupling. The $^{15}$NH$_3$ $J_K = 1_1-1_1$ consists of 13 hyperfine components, labelled with the quantum numbers $J$, $K$, $F$, and $I$. The $J_K = 2_2-2_2$ full hyperfine structure is made of 14 components, with the central 4 components merged in a single unresolved feature.
 \par
The spectroscopic fit to the \am lines is based on six free parameters: the local standard-of-rest velocity (\vlsr), the line velocity dispersion (\sigmav), the total column density $N_\mathrm{col}(\text{\am})$, the excitation and rotational temperature (\tex and $T_\mathrm{rot}$), and the ortho-to-para ratio. We fix the latter on 1 since we have access only to para states, even though we are aware that models predict $OPR =0.5 $ for \am \citep[see e.g.][fig 1.]{Sipila15}. Having detected two inversion transitions, both at high $\rm S/N$, allows us to constrain the remaining five parameters. 
\par
For the rarer isotopologue, we only detect the first (1,1) line. We cannot, therefore, constrain all the parameters since they become degenerate, also due to the fact that the hyperfine structure of the transition is very compact and assessment of the line optical depth is hence challenging. When fitting this transition, we keep the rotational and excitation temperature fixed to the values derived from the main isotopologues. Only \sigmav, \vlsr, and $N_\mathrm{col}(\text{\qam})$ are then fit. Assuming that the two isotopologues emit from the same region, one could also fix the kinematic parameters (\sigmav and \vlsr) of the rarer \qam to those of the main \am. However, we prefer to keep them free, in order to investigate the analogies and differences among them. We have however checked that the two approaches yield consistent results in terms of the isotopic ratio, within uncertainties. 
\par
{The assumption that $T_\mathrm{rot}$ is the same for the two isotopologues is not entirely correct, as the true constant quantity is the gas kinetic temperature $T_\mathrm{K}$. This assumption, however, introduces negligible errors. The relation between $T_\mathrm{rot}$ and $T_\mathrm{k}$ is derived from assuming a three-level system formed only by the states (1,1), (2,2), and (2,1); it depends on the energy difference between the (1,1) and the (2,2) levels (which differs by less than 1\% between \am and \qam) and on the ratio among the collisional coefficients $C_{2,2 \to 2,1}/C_{2,2 \to 1,1}$ \citep[cf.][]{Swift05}. The \qam collisional coefficients are not available. However, we expect that they can differ at most by 10-15\% (similar differences have been found in \nlow-bearing isotopologues of HNC and HCN, see \citealt{NavarroAlmaida23}). Furthermore, for $T_\mathrm{K}<15\, \rm K$, $T_\mathrm{rot} \approx T_\mathrm{K}$ with a maximum difference of $<10$\%.}
\par
The spectral fits yield, among the other parameters, the best-fit values and uncertainties for the column densities of the two isotopologues. The isotopic ratios are evaluated from the ratio of these column densities, and the associated uncertainties are computed using standard error propagation. 
\par
In order to produce the ammonia \ratio map in L1544 discussed in Sect.~\ref{sec:L1544_map}, we first selected the pixels where the integrated intensity of \qam (1,1) is above the $3\sigma$ level, and where the peak intensity of \am (2,2) is detected above the $5\sigma $ level. The latter constraint aims at improving the estimates of the rotational temperature and  We then follow the same procedure as for the single pointing described in Sect.~\ref{sec:L1544}. We fit the main isotopologue transitions simultaneously, obtaining ---among others--- the maps of the rotational temperature, the excitation temperature, and \am column density. The first two maps are used as fixed parameters to fit the \qam (1,1) transition. The ratio between the column density maps of the two isotopologues yields the nitrogen isotopic ratio map shown in Fig.~\ref{fig:L1544_map}. 
\par
{We highlight that the hypothesis that \tex is the same for both isotopologues could be untrue if they are not tracing the same regions in the source. Assessing \tex for the rarer isotopologue is not possible with the available data. We have, however, made a test, modifying the \tex values of \qam by $\pm 1\, \rm K$. The resulting isotopic ratios vary by $15-30$\%, which is comparable with the typical relative errors on \ratio. Furthermore, the comparison between the distinct sources and the trends within L1544 still hold. We conclude that our results do not depend strongly on this assumption.}
\section{Akaike information criterion\label{app:AIC_analysis}}
The Akaike information criterion determines if a model fit improves significantly when the number of free parameters is increased. To do so, the method computes the quantity
\begin{equation}
\text{AIC} = \chi^2+C+2n \; ,
\end{equation}
where $\chi^2$ is the chi-squared of the model, $n$ is the number of free parameters, and $C$ is a constant that depends, basically, on the noise level in the data. When fitting a spectrum with one or two velocity components, $C$ does not change, whilst the number of free parameters changes from $n=5$ to $n'=10$. The relevant quantity is hence the AIC variation:
\begin{equation}
\Delta_\mathrm{AIC} = \Delta \chi^2 +2 \times  (n'-n)  = \Delta \chi^2 - 10 \;. 
\end{equation}
The smaller the AIC value, the better the fit. The $ \chi^2 $ value for each fit is
\begin{equation}
 \chi^2  = N_\mathrm{ch}\times \frac{ \sigma_\mathrm{res}^2}{rms^2} \; , 
\end{equation}
where $ \sigma_\mathrm{res}$ is the standard deviation of the residuals of the fit, computed over the channels $ N_\mathrm{ch}$ where the model with two components is $> 10^{-6}$ (in order to keep $ N_\mathrm{ch}$ constant when varying the number of velocity components), and $rms = 9\, \rm mK$ is the noise, assumed to be constant in all channels.  The AIC variation we compute when increasing the number of velocity components is $\Delta_\mathrm{AIC}  = -19 \times 10^3$, hence the improvement of the fit is highly significant. 

\end{document}